\documentclass[twocolumn,times]{aastex62}

\newcommand{\kms}{km\,s$^{-1}$}
\newcommand{\ms}{m\,s$^{-1}$}

\newcommand{\yy}{YY~Gem}
\newcommand{\yya}{YY~Gem~A}
\newcommand{\yyb}{YY~Gem~B}

\newcommand{\fifps}[2]{\centering\resizebox{#1}{!}{\includegraphics{#2}}}

\received{\today}
\revised{\today}
\accepted{\today}

\shorttitle{Magnetic Field of YY Gem}
\shortauthors{Kochukhov \& Shulyak}

\begin{document}

\title{Magnetic Field of the Eclipsing M Dwarf Binary YY~Gem}

\correspondingauthor{Oleg Kochukhov}
\email{oleg.kochukhov@physics.uu.se}

\author[0000-0003-3061-4591]{Oleg Kochukhov}
\affil{Department of Physics and Astronomy, Uppsala University, Box 516, 75120 Uppsala, Sweden}

\author[0000-0001-8496-2482]{Denis Shulyak}
\affil{Max-Planck Institute for Solar System Research, Justus-von-Liebig-Weg 3, 37077 G\"ottingen, Germany}

\begin{abstract}
YY~Gem is a short-period eclipsing binary system containing two nearly identical, rapidly rotating, very active early-M dwarfs. This binary represents an important benchmark system for calibrating empirical relations between fundamental properties of low-mass stars and for testing theories of interior structure and evolution of these objects. Both components of YY~Gem exhibit inflated radii, which has been attributed to poorly understood magnetic activity effects. Despite a long history of magnetic activity studies of this system no direct magnetic field measurements have been made for it. Here we present a comprehensive characterisation of the surface magnetic field in both components of YY~Gem. We reconstructed the global field topologies with the help of a tomographic inversion technique applied to high-resolution spectropolarimetric data. This analysis revealed moderately complex global fields with a typical strength of 200--300~G and anti-aligned dipolar components. A complementary Zeeman intensification analysis of the disentangled intensity spectra showed that the total mean field strength reaches 3.2--3.4~kG in both components of YY~Gem. We used these results together with other recent magnetic field measurements of M dwarfs to investigate the relation between the global and small-scale fields in these stars. We also assessed predictions of competing magnetoconvection interior structure models developed for YY~Gem, finding that only one of them anticipated the surface field strength compatible with our observations. Results of our star spot mapping of YY~Gem do not support the alternative family of theoretical stellar models which attempts to explain the radii inflation by postulating a large spot filling factor. 
\end{abstract}


\section{Introduction}
\label{intro}

Detached eclipsing binary systems provide a unique possibility of a model independent determination of fundamental parameters of their stellar components \citep{torres:2010}. In particular, a combined analysis of the photometric observations of eclipses and radial velocity variation of individual components enables one to infer stellar radii and masses with a precision of a few per cent. This, in turn, permits comprehensive tests of predictions of the stellar evolution and model atmosphere theories. 

A striking result emerging from recent analyses of low-mass eclipsing binary systems is the finding that standard stellar evolutionary models underestimate the observed stellar radii by 5--15\% \citep[e.g.][]{torres:2002,ribas:2006,morales:2009,morales:2009a}. Many of the systems exhibiting this discrepancy are short-period binaries with rapidly rotating components. Such late-type stars are expected to host stronger magnetic fields as a result of enhanced efficiency of the rotationally-powered dynamo \citep{vidotto:2014,marsden:2014}. Therefore, it was proposed that magnetic activity associated with a rapid stellar rotation might be responsible for the observed radii inflation of late-type close binary stars \citep{chabrier:2007,lopez-morales:2007,morales:2008}.

Two types of effects are considered as plausible mechanisms of the interplay between magnetic field and stellar radii. On the one hand, formation of dark cool spots in the magnetised areas at the stellar surface suppresses radiative losses resulting in a radius increase compared to unspotted star of the same luminosity \citep{chabrier:2007,morales:2010,jackson:2014}. On the other hand, inflated radii can be explained by the modification of stellar structure by the magnetic suppression or stabilisation of the interior convection \citep{mullan:2001,feiden:2012a,feiden:2013,feiden:2014,macdonald:2013,macdonald:2014,macdonald:2017}. Relative importance of these two effects as a function of stellar mass as well as validity of alternative theoretical approaches to treating magnetoconvection in one-dimensional stellar structure models are hotly debated topics. Detailed observations and comprehensive modelling of selected benchmark binary systems is necessary to inform this debate and guide further theoretical development.

YY Geminorum (Castor C, HD\,60179 C, BD+32 1582, GJ 278 C) is a key object for understanding the impact of magnetic activity on the fundamental properties of low-mass stars. This is a close, double-line, eclipsing binary system with an orbital period of 0.81~d containing two nearly identical active dM1e stars. \yy\ is a part of the remarkable sextuplet system $\alpha$~Gem (Castor). Both Castor A and B are young, early-type spectroscopic binaries. The association to the $\alpha$~Gem group provides additional stringent constraints on the age and metallicity of \yy. The physical parameters of the \yy\ components were studied in detail by \citet{torres:2002}. These authors determined the mean stellar radius (0.6191~$R_\odot$) and mass (0.5992~$M_\odot$) with an accuracy of better than 1\% and demonstrated that these radii exceed theoretical predictions by 10--20\%. \yy\ is one of only four M dwarf eclipsing binaries with properties measured accurately enough to enable meaningful tests of stellar evolutionary models \citep{torres:2013}. It is thus extensively used as a reference object for empirical mass and radius calibrations \citep[e.g.][]{torres:2010,eker:2015,moya:2018}. A number of theoretical studies attempted to explain the inflated radii of the components of \yy\ by developing different versions of non-conventional stellar interior structure models which incorporated effects of magnetic field and surface inhomogeneities \citep{feiden:2013,macdonald:2014,macdonald:2017,jackson:2014}.

The ubiquitous manifestations of the magnetic activity of \yy, including frequent flaring, broad-band photometric and emission line variability, non-thermal X-ray and ultraviolet emission, have been investigated by many authors. Since the seminal work by \citet{kron:1952}, the out-of-eclipse photometric variation of \yy\ was interpreted in the context of the rotational modulation caused by a patchy, nonuniform surface brightness distribution on one or both components \citep{torres:2002,butler:2015}. Large and frequent flare events have been reported for the system and investigated with the help of multiwavelength monitoring campaigns spanning the frequency range from X-ray to radio \citep[][and references therein]{butler:2015}. \yy\ was also targeted by several comprehensive X-ray imaging and spectroscopic studies \citep{gudel:2001,stelzer:2002,hussain:2012}, which demonstrated that both components are extremely active and flaring frequently. Detailed magneto-hydrodynamical models of flares have been developed using \yy\ as a testbed \citep{gao:2008}.

Many previous studies emphasise the central role of magnetic field and related surface inhomogeneities for interpretation of the multitude of activity phenomena in the \yy\ system and for explaining anomalous radii of its binary components. However, literature contains little direct, quantitative information on the magnetic field properties and surface structure morphology of \yya\ and B. Several studies attempted to model the out-of-eclipse photometric time series observations of the combined light from the system in terms of dark surface spots \citep{torres:2002,butler:2015}. Such analyses generally require strong assumptions regarding the spot characteristics (e.g. spot shapes and spot-to-photosphere contrast have to be prescribed) and often struggle to constrain spot latitudes or unambiguously assign them to one or another component. A spectroscopic surface mapping with the help of the Doppler imaging (DI) method is potentially capable of providing a more reliable and higher-resolution information on the surface spot distributions, although it has its own caveats when applied to equator-on stars such as the components of \yy. Preliminary DI maps of both components of \yy\ were presented by \citet{hatzes:1995a}, but this work was not followed up with a detailed study. No direct measurements of the mean magnetic field modulus are available for \yy. The global magnetic field geometry of the components has not been characterised in detail either.

The main aim of our investigation of \yy\ is to supply observational constraints to magnetoconvection interior structure models of low-mass stars by performing a comprehensive characterisation of the surface magnetic field of both components. This requires a detailed spectroscopic and spectropolarimetric analysis of the system. As a by-product, this analysis yields revised estimates of the fundamental parameters of the components and enables us to study their inhomogeneous surface brightness distributions. The rest of this paper is organised as follows. Section~\ref{obs} describes observational material used in our study. The procedure of deriving the mean intensity and polarisation profiles is presented in Section~\ref{lsd}. Section~\ref{sbfit} discusses binary spectral disentangling and presents revised spectroscopic orbit of \yy. Section~\ref{zdi} briefly describes methodology of Doppler and Zeeman-Doppler imaging (ZDI) of double-line binary systems and presents results of application of this tomographic mapping technique to \yya\ and B. Finally, the summary of our main findings and discussion of their implications are given in Section~\ref{disc}.

\section{Observational Data}
\label{obs}

Our investigation of \yy\ is based on a set of archival high-resolution spectropolarimetric observations obtained with the ESPaDOnS spectropolarimeter at the Canada-France-Hawaii Telescope (CFHT). This instrument provides simultaneous coverage of the 370--1050~nm wavelength interval at the resolving power of $R=65,000$ and allows one to record either linear or circular polarisation spectra in addition to the usual intensity observations. \yy\ was observed in the circular polarisation mode, in the period from Jan 4 to Jan 17, 2012. With typically 3 observations per night, this campaign resulted in 36 individual circular polarisation observations. Each of these observations consisted of four 420~s sub-exposures, between which the polarimeter was reconfigured to exchange the optical path and  detector positions of the orthogonal polarisation beams. As discussed by \citet{bagnulo:2009}, this spatiotemporal polarimetric modulation technique is highly effective in removing spurious and instrumental polarisation features. The ESPaDOnS spectra were reduced by the UPENA pipeline running the {\sc Libre-ESpRIT} software \citep{donati:1997}.

We have used the CFHT Science Archive\footnote{\url{http://www.cadc-ccda.hia-iha.nrc-cnrc.gc.ca/en/cfht}} interface to retrieve 36 Stokes $V$ observations as well as 144 Stokes $I$ spectra corresponding to individual polarimetric sub-exposures. These data were reprocessed by improving continuum normalisation with the method described by \citet{rosen:2018}. In addition, we calculated heliocentric Julian dates (HJDs) of mid-exposures from the time-stamps provided in the ESPaDOnS data files using Astrolib\footnote{\url{https://idlastro.gsfc.nasa.gov}} IDL routines.

According to the diagnostic information available from the headers of the pipeline-reduced ESPaDOnS spectra, the median signal-to-noise ratio (SNR) of the Stokes $I$ spectra of \yy\ is 280 per extracted spectrum pixel at $\lambda$\,=\,870~nm. The random photon noise of the circular polarisation spectra is characterised by the median SNR of 510 relative to the continuum of the corresponding Stokes $I$ spectra. Columns 1 and 3 of Table~\ref{tbl:obs} report individual HJD and SNR values, respectively, for all 144 Stokes $I$ spectra analysed here. The set of 36 Stokes $V$ observations discussed in our paper corresponds to groups of four consecutive entries in Table~\ref{tbl:obs}. 

It should be noted that, while the 420~s exposure time of individual sub-exposures is negligible compared to the 0.81~d orbital period of \yy, the total time ($\approx$\,1805~s) of acquiring four such sub-exposures comprising a single Stokes $V$ observation corresponds to 2.6\% of the orbital cycle. The radial velocities of the \yy\ components can change by up to $\approx$\,25~\kms\ during this time interval. It is therefore essential to account for the resulting orbital radial velocity smearing in any analysis of the Stokes $V$ data.

\begin{table}[!h]
\caption{Radial velocity measurements of the \yy\ components. \label{tbl:obs}}
{\small
\begin{tabular}{lllccc}
\hline
\hline
Reduced HJD & Phase & SNR & $V_{\rm A}$ & $V_{\rm B}$ & Orb. \\
& & & (\kms) & (\kms) & fit \\
\hline
  55930.8055 &  0.723 &  309 &  $122.31$ & $-116.84$ & y \\
  55930.8109 &  0.729 &  313 &  $123.10$ & $-117.77$ & y \\
  55930.8162 &  0.736 &  305 &  $123.67$ & $-118.23$ & y \\
  55930.8215 &  0.743 &  306 &  $124.05$ & $-118.59$ & y \\
  55930.9408 &  0.889 &  297 &  $ 79.95$ &  $-76.95$ & y \\
  55930.9462 &  0.896 &  289 &  $ 76.07$ &  $-73.05$ & y \\
  55930.9515 &  0.902 &  286 &  $ 72.03$ &  $-68.99$ & y \\
  55930.9569 &  0.909 &  282 &  $ 68.30$ &  $-64.19$ & y \\
\hline
\end{tabular}
}
\tablecomments{Table~\ref{tbl:obs} is published in its entirety in the electronic edition of the \textit{Astrophysical Journal}. A portion is shown here for guidance regarding its form and content.
}
\end{table}

\section{Least-squares Deconvolved Profiles}
\label{lsd}

Similar to most other cool active stars, circular polarisation signatures are too weak to be detected in individual spectral lines of \yy. In this situation a multi-line polarisation diagnostic technique is required to derive high-quality average Stokes $V$ profiles suitable for reliable magnetic field detection and detailed modelling of the global surface field topology. Calculation of the mean intensity profiles also provides input spectra for precise radial velocity measurements of binary star components. 

In this study we  employed the least-squares deconvolution \citep[LSD,][]{donati:1997} technique in its implementation by \citet{kochukhov:2010a}. This multi-line method approximates stellar spectra as a superposition of spectral lines represented by a scaled and shifted mean profile. Using this representation, equivalent to convolution of a mean profile and a line mask in the velocity space, one can compute mean intensity or polarisation spectra with a series of matrix operations. The input data necessary for this calculation consist of an observed spectrum, corresponding error bars, and a line mask containing information on the line positions and relative weights (line depth for Stokes $I$ and the product of line depth, wavelength, and the effective Land\'e factor for Stokes $V$).

\begin{figure}[!t]
\fifps{8.55cm}{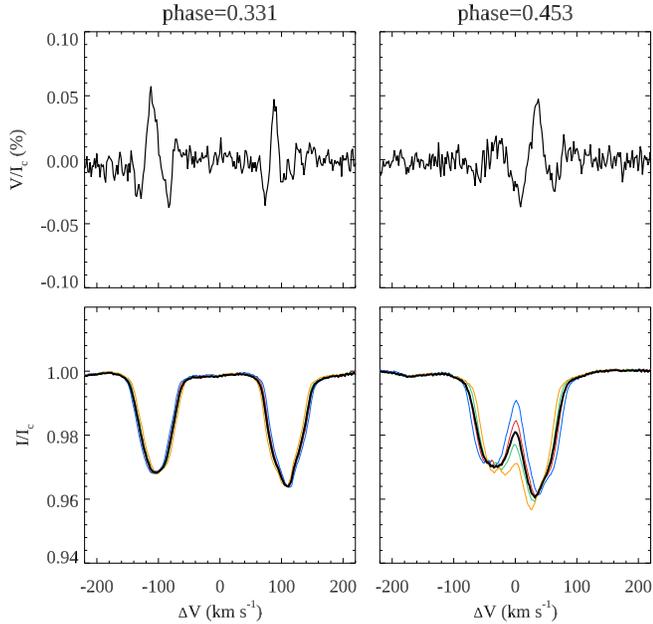}
\caption{
Representative Stokes $V$ (upper row) and Stokes $I$ (lower row) LSD profiles of \yy. The two columns show spectra for the orbital phase 0.331 (left) and 0.453 (right). The thin colour lines in the lower panels illustrate the Stokes $I$ profiles corresponding to four spectropolarimetric sub-exposures.
}
\label{fig:lsd_one}
\end{figure}

We used the VALD3 database\footnote{\url{http://vald.astro.uu.se}} \citep{ryabchikova:2015} together with the solar metallicity MARCS \citep{gustafsson:2008} model atmosphere with $T_{\rm eff}$\,=\,3800~K and $\log g$\,=\,4.5 \citep{torres:2002} to compile an absorption line list appropriate for the \yy\ components. The final LSD line mask, comprising 5059 atomic lines, was derived by excluding spectral regions affected by broad stellar features or telluric absorption and then selecting lines with the residual depth greater than 0.2 of the continuum. The polarisation line weights were normalised according to the mean wavelength $\lambda_0$\,=\,636~nm and the mean effective Land\'e factor $z_0$\,=\,1.22.

The LSD line-averaging procedure was applied separately to the 144 Stokes $I$ spectra corresponding to individual polarimetric sub-exposures and to the Stokes $I$ and $V$ spectra comprising 36 complete circular polarisation observations. Representative Stokes $I$ and $V$ LSD profiles are illustrated in Fig.~\ref{fig:lsd_one} for two orbital phases. This figure also shows the Stokes $I$ LSD spectra corresponding to groups of four consecutive spectropolarimetric sub-exposures. It is evident that these spectra exhibit non-negligible radial velocity shifts due to the orbital motion of the \yy\ components. It is also apparent that the equivalent width of the primary LSD profile (shifted to negative velocities in Fig.~\ref{fig:lsd_one}) is systematically smaller compared to the profile of \yy~B.

Thanks to co-adding information from thousands of individual lines, the Stokes $V$ LSD spectra of \yy\ boast a SNR of $\approx$\,15,000 per 1.8~\kms\ velocity bin, which corresponds to a factor of 30 gain relative to the polarimetric sensitivity of the original polarisation spectra. Variable polarisation signatures with a typical semi-amplitude of $5\times10^{-4}$ are detected for 34 out of 36 observations with a typical significance (full amplitude of the signal divided by the error bar) of 12$\sigma$. The two Stokes $V$ spectra yielding no detection have a greatly inferior quality (SNR of only about 1300 in the LSD Stokes $V$ profiles) compared to the rest of the data and are excluded from subsequent analysis.

\section{Spectral Disentangling and Orbital Solution}
\label{sbfit}

\begin{figure*}[!th]
\fifps{8.2cm}{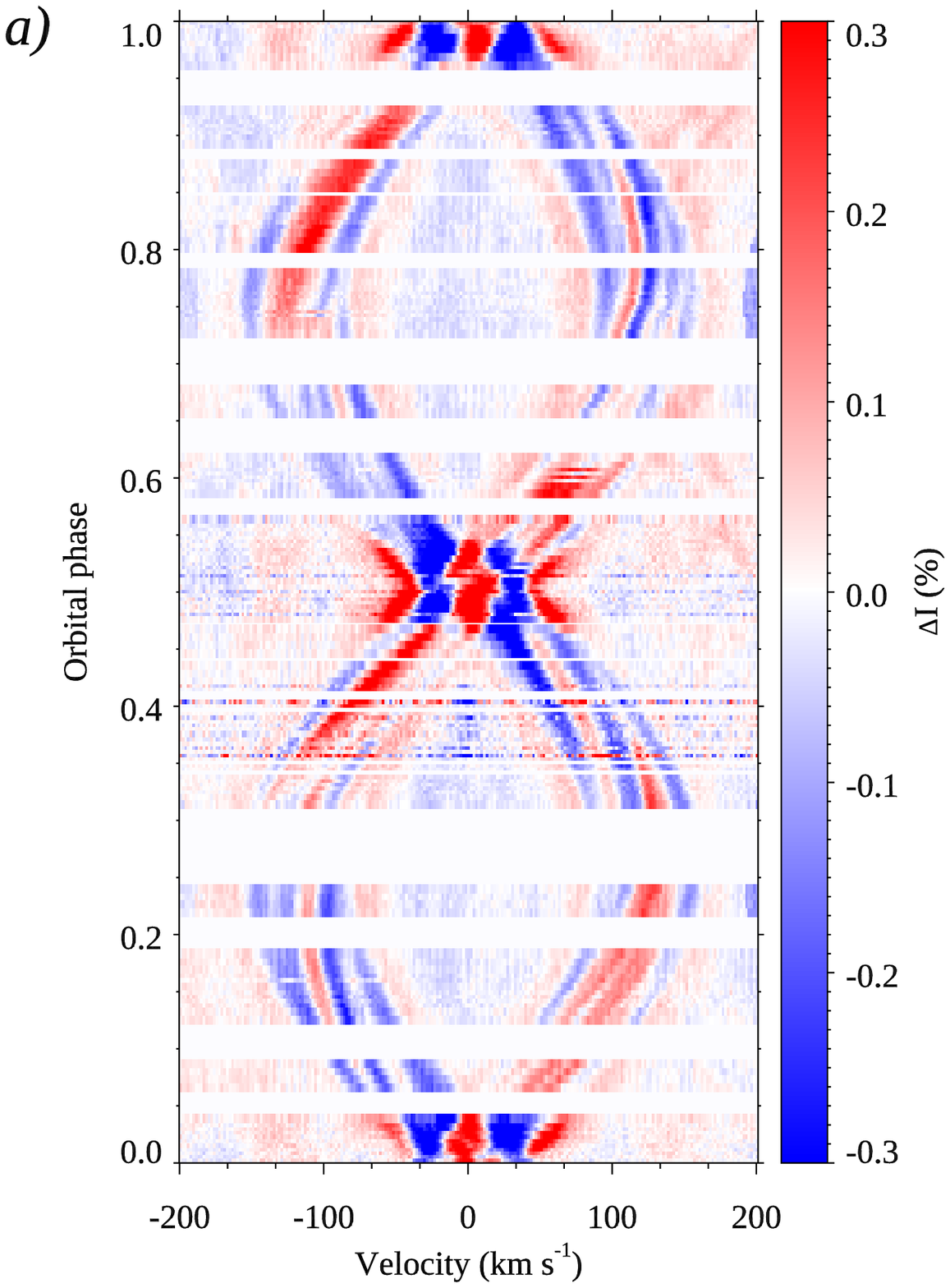}\hspace*{0.5cm}
\fifps{8.2cm}{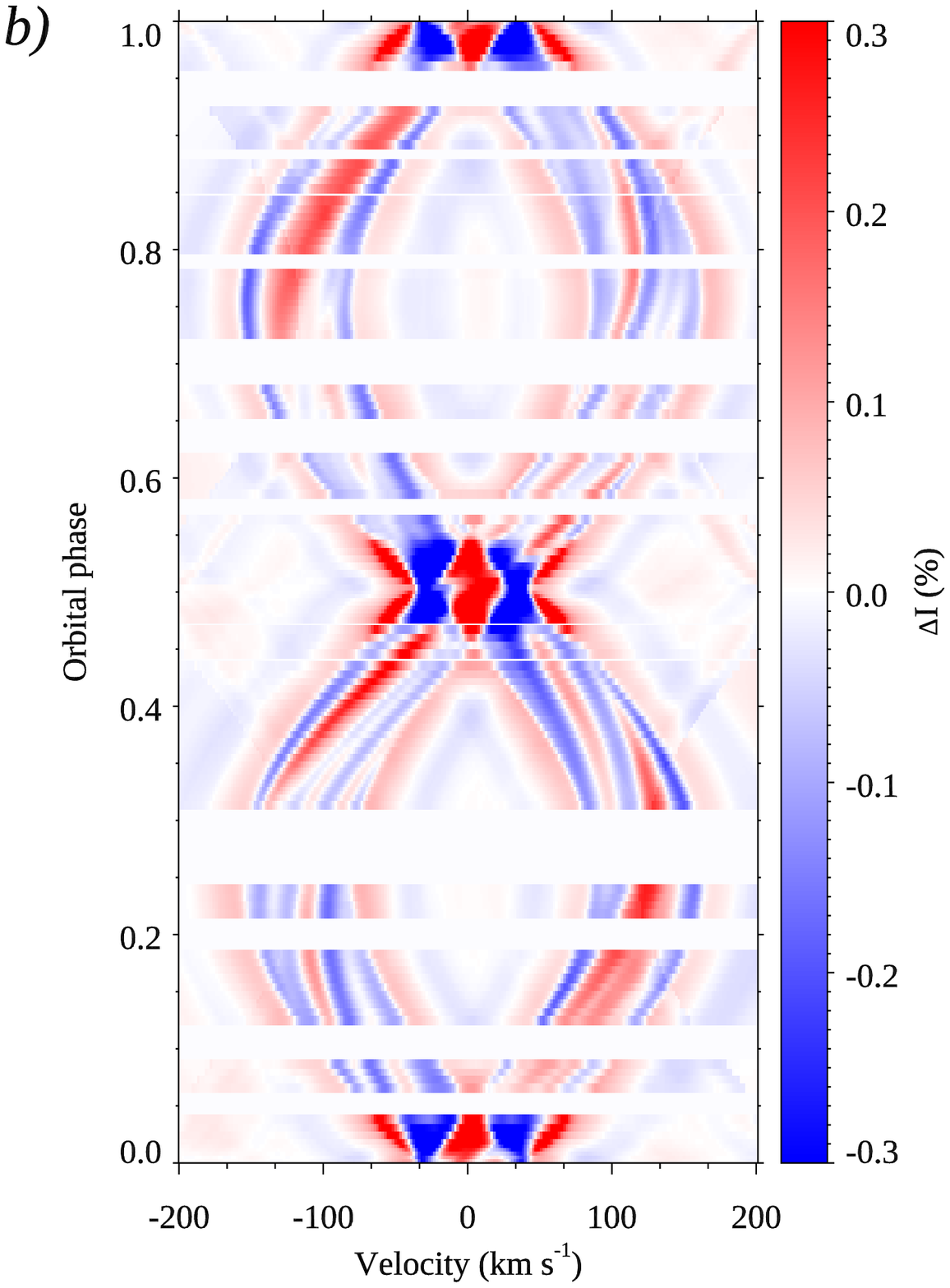}
\caption{
Dynamic residual Stokes $I$ LSD profiles of \yy. The observed (a) and model (b) residual spectra are plotted as a function of the orbital phase. The colour scale corresponds to $\pm0.3$\% intensity range.
}
\label{fig:dynamic}
\end{figure*}

The spectroscopic binary nature of \yy\ significantly complicates analysis of individual components due to complex, time-dependent line blending seen in the observed spectra. Nevertheless, redundant information contained in the observations collected at different orbital phases can be exploited to derive accurate radial velocities of the components, obtain their high-quality separated spectra and study line profile variability. To this end, we applied different spectral disentangling methods to LSD profiles and to selected regions of Stokes $I$ spectra.

First, 144 Stokes $I$ LSD spectra were analysed with the disentangling code described by \citet{folsom:2010} and applied by several other studies \citep[e.g.][]{rosen:2018,kochukhov:2018b}. This procedure performs an iterative derivation of the binary component velocities and mean profile shapes using a weighted least-squares fit of a set of observed LSD profiles. The method assumes that observations at each orbital phase can be represented by a superposition of two constant spectra with variable radial velocity shifts. The disentangling calculation starts with an initial guess of radial velocities of the components, for which we adopted the orbital solution by \citet{torres:2002}. Mean stellar spectra are then derived with a least-squares optimisation algorithm by fitting observations at all available orbital phases. In the next step the individual radial velocities are refined with another least-squares fit keeping the mean spectra fixed. This two-step procedure was repeated, 7 times in this case, until the convergence was achieved for all radial velocities and all spectral bins of the disentangled spectra.

The basic assumption of the disentangling procedure that the intrinsic stellar spectra are constant is not fulfilled during the primary and secondary eclipses. Consequently, our radial velocity estimates are less accurate during that part of the orbit. In addition, the LSD Stokes $I$ profiles of both components of \yy\ are affected by surface spots at all orbital phases. Nevertheless, as demonstrated by \citet{rosen:2018}, our binary spectral disentangling procedure is robust against such profile distortions.

The LSD profile disentangling is illustrated in Fig.~\ref{fig:dynamic}a, which shows dynamic residual Stokes $I$ spectrum obtained by subtracting the model two-component spectra from the observed profiles. This plot enables us to assess the intrinsic spectral variability of the components of \yy. It is clear that both stars show variability outside eclipses. With an amplitude of few times $10^{-3}$, this variability is relatively weak and is not detectable in typical individual spectral lines. The gradually evolving profile distortions seen in Fig.~\ref{fig:dynamic}a reveal surface inhomogeneities, which rotate in and out of view. The profile variability pattern of both components is perfectly phased with the orbital period, implying a synchronous rotation with $P_{\rm rot}^{\rm A}=P_{\rm rot}^{\rm B}=P_{\rm orb}$. There is no evidence of the surface structure evolution within the 13 d time span covered by the observations. The spot signatures encompass the entire width of the line profiles, suggesting that the surface structures responsible for this behaviour are located preferentially at low latitudes. One should bear in mind, however, that this residual profile analysis is not sensitive to any axisymmetric surface features, such as polar spots.

\begin{table}[!t]
\caption{Spectroscopic orbital solution for \yy. \label{tbl:orbit}}
\begin{tabular}{ll}
\hline
\hline
Parameter & Value  \\
\hline
\multicolumn{2}{l}{Fitted quantities:} \\
$P_{\rm orb}$ (d) & 0.814282212\tablenotemark{a} \\
$HJD_0$ & $2449345.116643\pm0.000068$ \\
$K_{\rm A}$ (\kms) & $121.337\pm0.072$ \\
$K_{\rm B}$ (\kms) & $121.264\pm0.064$ \\
$\gamma$ (\kms) & $2.287\pm0.038$ \\
$e$ & 0.0\tablenotemark{b} \\
\multicolumn{2}{l}{Derived quantities:} \\
$M_{\rm B}/M_{\rm A}$ & $1.00060\pm0.00080$ \\
$M_{\rm A} \sin^3 i$ ($M_\odot$) &  $0.60217\pm0.00058$ \\
$M_{\rm B} \sin^3 i$ ($M_\odot$) &  $0.60253\pm0.00060$ \\
$M_{\rm A}$  ($M_\odot$) &  $0.60597\pm0.00058$\tablenotemark{c} \\
$M_{\rm B}$  ($M_\odot$) &  $0.60633\pm0.00061$\tablenotemark{c} \\
$a_{\rm A} \sin i$ ($\times 10^6$ km) & $1.35863\pm0.00081$ \\
$a_{\rm B} \sin i$ ($\times 10^6$ km) & $1.35782\pm0.00072$ \\
\hline
\end{tabular}
\tablenotetext{a}{Adopted from \citet{torres:2002}.}
\tablenotetext{b}{Circular orbit is assumed.}
\tablenotetext{c}{Calculated assuming $i=86\fdg29\pm0\fdg1$ \citep{torres:2002}.}
\end{table}

The radial velocities of \yy\ A and B\footnote{Following the naming convention of previous publications, we refer to the star eclipsed at phase 0.0 as the ``primary'' or ``component A''.} obtained with the help of Stokes $I$ LSD profile disentangling are listed in columns 4 and 5 of Table~\ref{tbl:obs}.  Using these data we redetermined the spectroscopic orbital parameters of \yy\ adopting a fixed orbital period of $P_{\rm orb}$\,=\,0.814282212~d \citep{torres:2002}. The orbital parameters were optimised with a non-linear least-squares fit in IDL using the {\sc Mpfit} package \citep{markwardt:2009} and the {\tt helio\_rv} Astrolib routine. 48 radial velocity measurements close to phases 0.0 and 0.5 were excluded from the fit, as indicated by entries in the last column of Table~\ref{tbl:obs}.

\begin{figure}[!t]
\fifps{8.55cm}{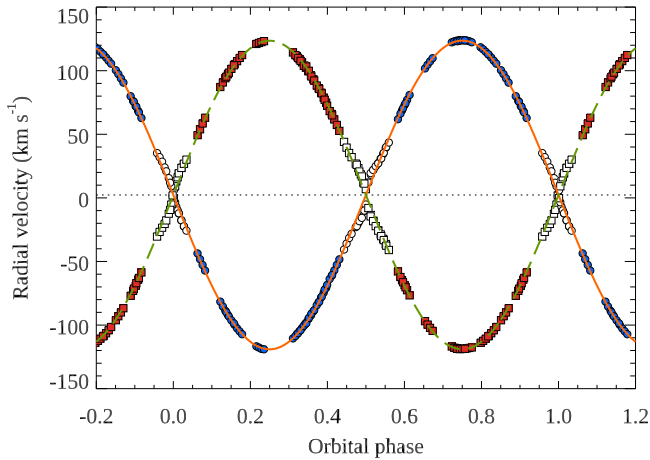}
\caption{
Orbital radial velocity variation of the \yy\ components. The symbols show measurements for the primary (circles) and secondary (squares). The solid and dashed lines illustrate the orbital solution constrained by the radial velocity measurements outside eclipses (filled symbols). Measurements at the eclipse phases (open symbols) are excluded from the fit.
}
\label{fig:orbit}
\end{figure}

\begin{figure*}[!th]
\fifps{16cm}{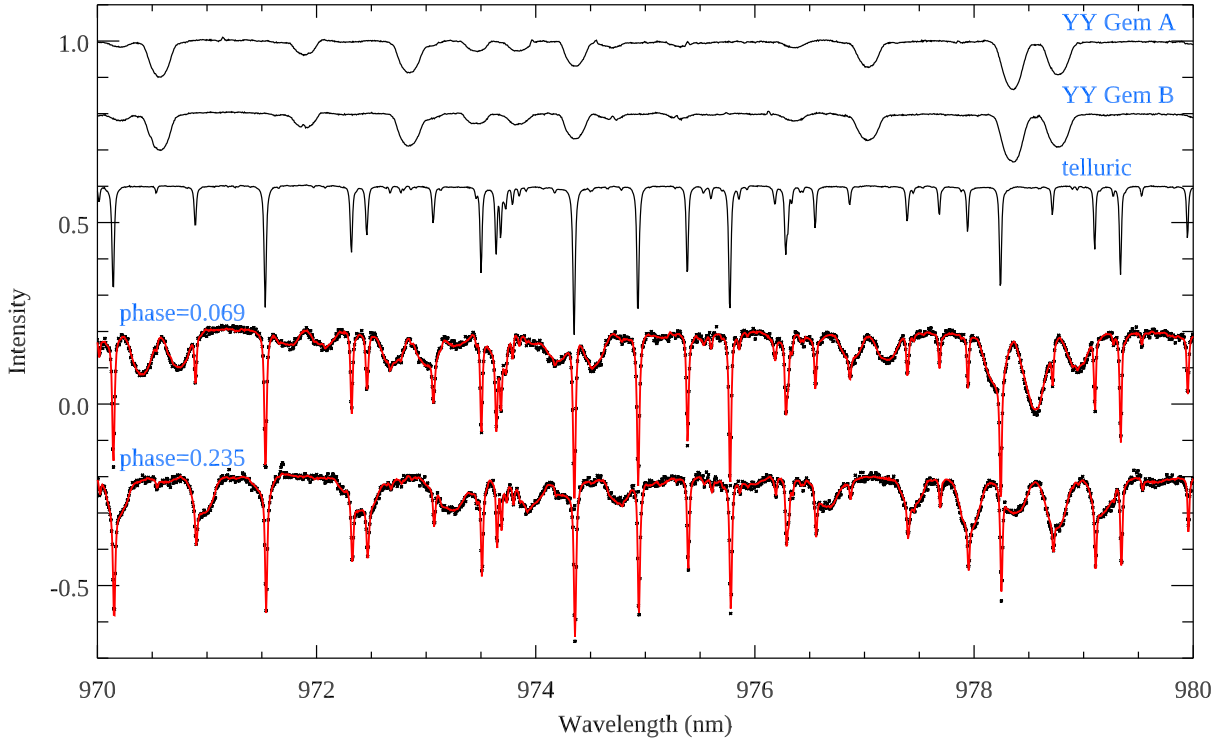}
\caption{
Illustration of the three-component disentangling of \yy\ spectra. The top three curves show the disentangled, time-averaged spectra of \yy\ A and B as well as the mean telluric absorption spectrum. The bottom curves compare observations (symbols) at two representative orbital phases with the composite model spectra (solid curve) including the stellar and telluric contributions. Spectra are offset vertically for display purposes.
}
\label{fig:s3fit}
\end{figure*}

The observed radial velocities of the \yy\ components are compared to the orbital fit in Fig.~\ref{fig:orbit}. The standard deviation of this fit is 0.46~\kms\ for \yya\ and 0.65~\kms\ for \yyb, which represents a factor of 5 to 7 improvement relative to the radial velocity analysis by \citet{torres:2002}. The revised orbital parameters are reported in Table~\ref{tbl:orbit}. In agreement with previous studies of \yy, we found the orbit to be circular ($e=0.00035\pm0.00028$) and therefore fixed eccentricity to zero in the final orbital fit. The error bars listed in the table are formal uncertainties calculated by {\sc Mpfit} and result by assigning uncertainties to the RV measurements equal to the standard deviations of the RVs around the best fit quoted above. This procedure results in a reduced chi-square of one.

Table~\ref{tbl:orbit} provides our estimates of the mass ratio $M_{\rm B}/M_{\rm A}$, scaled masses $M_{\rm A,B} \sin^3 i$, and semi-major axes $a_{\rm A,B} \sin i$. Finally, we report the component masses which follow from our spectroscopic orbital elements and the orbital inclination angle $i=86\fdg29\pm0\fdg1$ \citep{torres:2002}. The orbital phases of 144 Stokes $I$ observations calculated with the updated ephemeris are given in the second column of Table~\ref{tbl:obs}.

Another version of the spectral disentangling calculation was applied to the 963--985~nm wavelength region in order to obtain mean component spectra for the magnetic intensification analysis presented below. In this case our goal was to infer the component spectra using previously determined orbital velocities, which simplifies disentangling. On the other hand, the presence of strong, variable telluric absorption in this wavelength region adds a significant complication. Rather than using an observed telluric standard star or theoretical telluric spectrum we implemented telluric correction as part of the modified spectral disentangling analysis. The observations at each orbital phase were represented with a superposition of the two stellar components plus a telluric contribution. The latter spectrum was assumed to be shifted according to the (known) heliocentric radial velocity correction and scaled for individual observations following the usual power-law relation \citep[e.g.][]{cotton:2014}. This composite spectral model was iteratively fitted to all Stokes $I$ observations simultaneously, yielding telluric-corrected, disentangled time-averaged spectra of \yy\ A and B with a SNR in excess of 500. This three-component disentangling procedure is illustrated in Fig.~\ref{fig:s3fit}, which shows the derived mean stellar spectra, the average telluric spectrum and the fit to observations for two representative orbital phases. 

\section{Zeeman-Doppler imaging}
\label{zdi}

Doppler and Zeeman-Doppler imaging are powerful inversion techniques for indirect studies of surface structure of cool active stars \citep{kochukhov:2016}. Here we carried out a tomographic reconstruction of the surface brightness and magnetic field maps of both components of \yy\ using the {\sc InversLSDB} binary Zeeman-Doppler imaging code described by \citet{rosen:2018}. This indirect imaging method, developed from the {\sc InversLSD} code \citep{kochukhov:2014}, enables mapping of inhomogeneities on the surfaces of one or both components of a spectroscopic binary system using LSD profile intensity and/or polarisation observations. The forward spectral modelling implemented in {\sc InversLSDB} accounts for eclipses and can be performed either assuming spherical stellar shapes, an arbitrary eccentric binary orbit, arbitrary inclinations and rotation periods of the components (typical of wide misaligned binaries) or using the Roche-lobe geometry to describe detached or contact co-rotating binary components with aligned orbital and rotational axes (typical of close binaries). In the present analysis of \yy\ we used the latter Roche-lobe geometry mode. This treatment, fully appropriate for this system, implies that the rotation of the components is synchronised with the orbital motion, the stars rotate as solid bodies, and their, generally non-spherical, shapes are are given by equipotential surfaces. A set of system parameters necessary for binary spectrum synthesis calculations with {\sc InversLSDB} includes the component masses, the orbital period and inclination, and the two values of Roche-surface potentials or, equivalently, stellar polar radii. In addition, one has to specify relative surface brightness of the components to reproduce the observed line depth ratio.

\subsection{Brightness Distribution}
\label{mapi}

In the first step of tomographic analysis of \yy\ we modelled the Stokes $I$ LSD profiles with the goal to determine several nuisance parameters and derive stellar brightness distributions. We adopted the orbital parameters and stellar masses according to the results of Sect.~\ref{sbfit}. The orbital inclination $i=86\fdg29$ was taken from \citet{torres:2002}. 

The synthetic LSD profile calculations were based upon the Unno-Rachkovsky \citep{polarization:2004} analytical local line profile model. We used the mean line parameters reported in Sect.~\ref{lsd} and adjusted the local profile width to match the theoretical disk-centre LSD profile obtained from the {\sc Synth3} \citep{kochukhov:2007d} intensity spectrum calculated with the $T_{\rm eff}$\,=\,3800~K and $\log g$\,=\,4.5 MARCS model atmosphere. Radiative transfer calculations with the same code were used to establish the centre-to-limb variation of the continuum intensity at $\lambda$\,=\,636~nm. We found that the square root limb-darkening law \citep[e.g.][]{claret:2017} with the coefficients $c=0.115$ and $d=0.757$ provides the best representation of the model atmosphere predictions.

\begin{figure*}[!th]
\fifps{13cm}{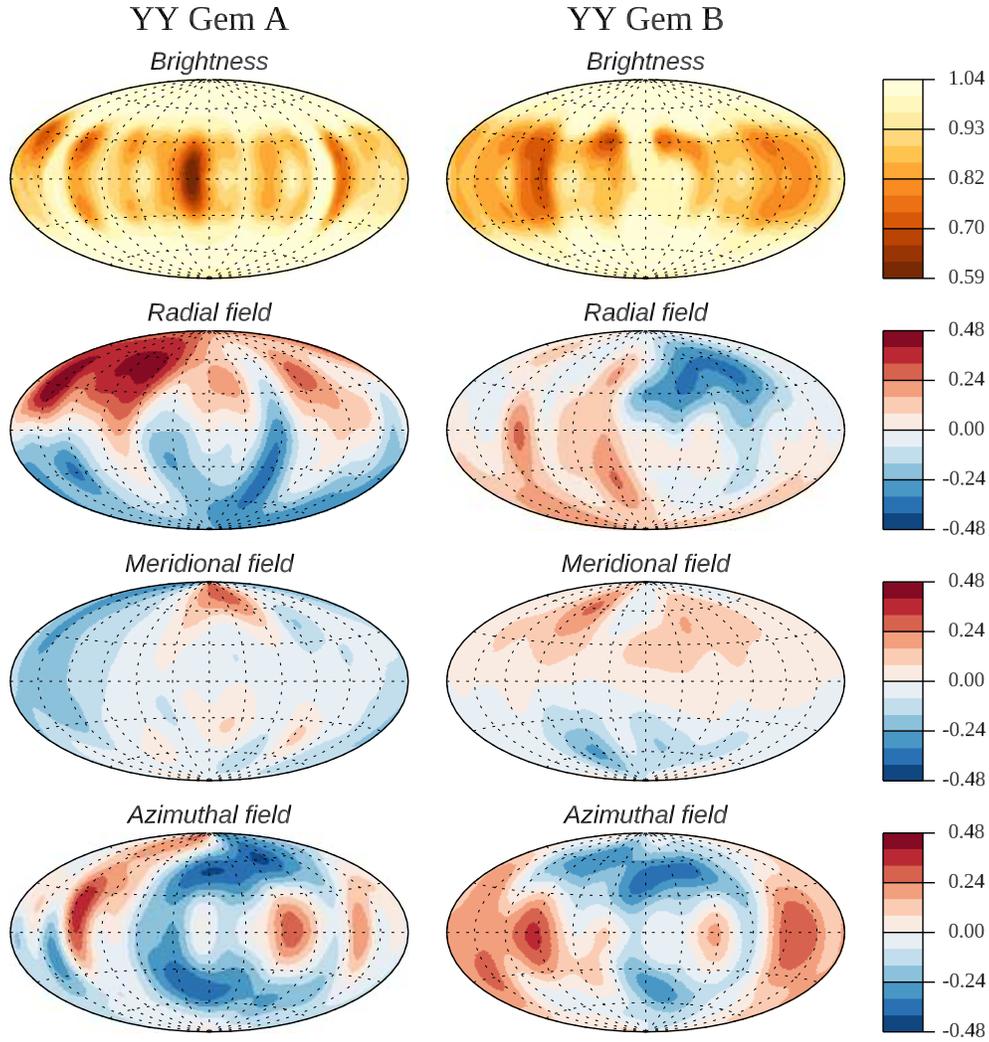}
\caption{
Brightness and magnetic field maps of \yya\ (left column) and \yyb\ (right column). The surface images are displayed in the Hammer-Aitoff projection, with the central meridian corresponding to 180\degr\ longitude. The topmost row shows the brightness distributions obtained from the Stokes $I$ profiles. The second to fourth rows show maps of the radial, meridional, and azimuthal field components derived from the Stokes $V$ spectra. The side colour bars indicate the relative continuum brightness and the field strength in kG. 
}
\label{fig:zdi_maps}
\end{figure*}

The equivalent width of the local profile, the relative surface brightness, and the radii of the components were optimised by fitting observations. As a result of this analysis we established that the best description of the LSD Stokes $I$ profiles of \yy\ is obtained with a surface brightness ratio of $I_{\rm A}/I_{\rm B}=0.88\pm0.01$ and equivalent volume radii of $R_{\rm A}=0.614\pm0.010$~$R_\odot$ and $R_{\rm B}=0.612\pm0.010$~$R_\odot$. 
The maximum difference between the radii measured along ($R_{\rm in}$) and perpendicular ($R_{\rm pole}$) to the axis connecting the centres of masses of the two components is about 1\%, implying that \yy\ A and B exhibit a negligible deviation from the spherical shapes.

The unequal surface brightness of the two components introduced above is necessary to reproduce the observed systematically weaker LSD Stokes $I$ profiles of \yy~A compared to \yy~B (see Fig.~\ref{fig:lsd_one}). This line depth difference can be also reproduced by directly postulating a different equivalent width of the local profiles, leading to the same modelling results. Here we treat $I_{\rm A}/I_{\rm B}$ as a nuisance parameter, without attempting to investigate its physical meaning. Nevertheless, this systematic difference of the line strengths for the two stars with essentially identical mass and radius is somewhat surprising and may indicate a $T_{\rm eff}$ difference of $\approx$\,90~K.

The final brightness maps derived for the components of \yy\ are presented in the upper panels of Fig.~\ref{fig:zdi_maps}. These continuum brightness distributions were derived using the modified Tikhonov regularisation method \citep[see][]{hackman:2016,rosen:2018}, which numerically stabilises the surface imaging problem by minimising the local map contrast and limiting deviation from the default brightness level of the unspotted photosphere. The surface maps in Fig.~\ref{fig:zdi_maps} are shown in the Hammer-Aitoff projection. The stellar longitude is counted counter-clockwise, in the same direction as the stellar rotation, and increases from left to right in the figure. The central meridian corresponds to the longitude of 180\degr.

Owing to a large inclination angle, the Doppler mapping method is unable to distinguish the Northern and Southern stellar hemispheres, except when the surface features are partially obscured during the primary and secondary eclipses. This is why the recovered dark spot geometries are symmetric with respect to the stellar equators everywhere except close to 0\degr\ longitude (phase 0.0) for \yya\ and 180\degr\ longitude (phase 0.5) for \yyb. The overall continuum brightness contrast required to reproduce observations is relatively small. The darkest feature ($\approx$\,40\% intensity contrast) is found at the surface of \yya\ very close to the point facing the secondary. \yyb\ shows no corresponding dark spot. Apart from this difference, the components of \yy\ have statistically similar degree of spot coverage. In agreement with the qualitative Stokes $I$ profile analysis in Sect.~\ref{sbfit}, dark spots are found only within $\pm$30\degr\ latitude band. Neither star exhibits a polar spot, which agrees with the findings by \citet{hatzes:1995a}.

Given the large number of observations and a small amplitude of the Stokes $I$ profile variability it is impractical to document DI analysis with the traditional comparison of the observed and calculated line profiles. Instead, we produced a dynamic residual profile plot similar to Fig.~\ref{fig:dynamic}a, but using the model LSD profiles as an input for disentangling procedure. The result is shown in Fig.~\ref{fig:dynamic}b. The observed residual profile variability pattern is successfully reproduced by the DI binary star model. The out of eclipse profile distortions, tracing the brightness inhomogeneities, have the same shape and amplitude as in observations. The largest residuals occur at the eclipse phases and have identical structure in both observations and the model, confirming that these residuals originate from the intrinsic limitation of the standard spectral disentangling treatment at these orbital phases.

\subsection{Global Magnetic Field Topology}
\label{mapb}

\begin{figure}[!th]
\fifps{8.55cm}{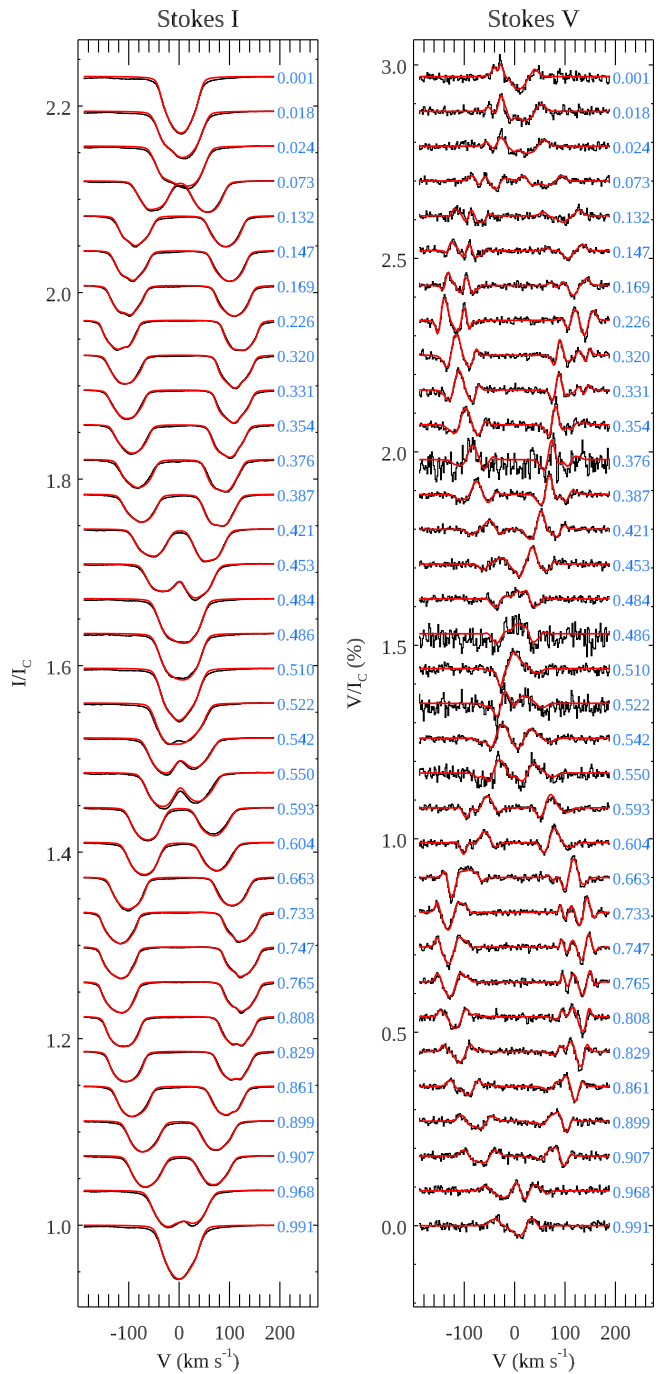}
\caption{
ZDI fit (solid lines) to the observed (histograms) Stokes $I$ and $V$ LSD profiles of \yy. The spectra corresponding to different orbital phases are offset vertically. The phases are indicated to the right of each profile.
}
\label{fig:zdi_prf}
\end{figure}

We applied ZDI analysis to infer the global magnetic field topologies of the \yy\ components from LSD observations of the circular polarisation signatures in metal line profiles. This observable provides detailed information on the geometrical structure of the vector field but is sensitive only to a large-scale, organised magnetic field component. Due to cancellation of polarisation signals coming from surface regions with opposite field polarities small-scale fields do not contribute to the disk-integrated stellar polarisation spectra. Therefore, ZDI maps derived here do not inform us about the total magnetic field energy and do not provide a realistic assessment of the total mean field modulus.

With these caveats in mind, we have carried out a tomographic reconstruction of the global field topologies of \yya\ and B following the ZDI methodology discussed by \citet{kochukhov:2014} for single stars and by \citet{rosen:2018} for binaries. The vector surface field distribution of each star is parametrised in terms of a general spherical harmonic expansion. Three sets of spherical harmonic coefficients are employed to describe the radial poloidal, the horizontal poloidal, and the horizontal toroidal fields. Each set is comprised of the spherical harmonic terms with all possible azimuthal numbers $m$ and the angular degrees $\ell$ from $\ell=1$ to some maximum value $\ell=\ell_{\rm max}$. In this study we have chosen $\ell_{\rm max}=10$, resulting in 360 spherical harmonic coefficients for each star. All these coefficients are treated as free parameters and are adjusted simultaneously by fitting 34 observed Stokes $V$ LSD profiles, starting from the zero-field initial guess. The inhomogeneous brightness distribution derived in the previous step of Stokes $I$ DI analysis is taken into account in the calculation of synthetic circular polarisation spectra.

The line profile synthesis relies on the Unno-Rachkovsky formulas, assuming that the response of the local LSD profile to magnetic field is equivalent to that of a normal Zeeman triplet with the effective Land\'e factor $z$\,=\,1.22. As mentioned in Sect.~\ref{obs}, ESPaDOnS polarisation observations are constructed from sets of four consecutive sub-exposures obtained over a non-negligible fraction of the orbital period. We explicitly accounted for the corresponding rotational and orbital phase smearing by calculating each synthetic profile with a 5-point trapezoidal integration over an appropriate orbital phase interval.

\begin{table}[!t]
\caption{Magnetic field characteristics of \yy. \label{tbl:mag}}
\begin{tabular}{lcc}
\hline
\hline
Parameter & \yya\ & \yyb\  \\
\hline
\multicolumn{3}{l}{From ZDI analysis:} \\
$\langle B_V \rangle$ (kG) & 0.260 & 0.205 \\
$\langle |B_{\rm r}| \rangle$ (kG) & 0.168 & 0.098 \\
$\langle B_{\rm h} \rangle$ (kG) & 0.179 & 0.162 \\
$E_{\rm pol}$\tablenotemark{a} (\%) & 70.7 & 71.5 \\
$E_{m<\ell/2}$\tablenotemark{b} (\%) & 58.1 & 44.8 \\
$E_{\ell=1}$\tablenotemark{c} (\%) & 52.4 & 45.5 \\
\multicolumn{3}{l}{From Zeeman intensification analysis:} \\
$\langle B_I \rangle$ (kG) & 3.44 & 3.15 \\
$B$ (kG) & 4.60 & 3.65 \\
$f$ & 0.75 & 0.86 \\ 
\hline
\end{tabular}
\tablenotetext{a}{Fractional energy of the poloidal field component.}
\tablenotetext{b}{Fractional energy of the axisymmetric field component.}
\tablenotetext{c}{Fractional energy of the dipolar field component.}
\end{table}

The magnetic inversion problem is numerically stabilised with the help of the harmonic regularisation method \citep{kochukhov:2014}. Specifically, the inversion is constrained by a penalty function, which minimises the total magnetic field energy of the ZDI map and restricts contributions of higher-order harmonic terms. The optimal regularisation parameter is determined following the procedure described by \citet{kochukhov:2017}.

The global magnetic field maps of the \yy\ components are shown in Fig.~\ref{fig:zdi_maps}. Table~\ref{tbl:mag} reports numerical characteristics of these field geometries, which might be of interest in the context of comparison to ZDI results obtained for other active stars. The final Stokes $V$ profile fit provided by the ZDI model is illustrated in Fig.~\ref{fig:zdi_prf}. The reduced $\chi^2$ of this fit is 0.91, implying that all significant details of the observed polarisation signatures are successfully reproduced by the model spectra.

We found that the primary has a somewhat stronger global magnetic field, with the mean field strength of 260~G compared to 205~G for the secondary. The mean unsigned radial magnetic field \citep[e.g.][]{vidotto:2014} is 168~G and 98~G for the primary and secondary, respectively. The maximum surface field strength is 546~G for \yya\ and 515~G for \yyb. Both stars possess moderately complex global field configurations, dominated by dipolar components (45--52\% of the magnetic field energy is contained in $\ell=1$ modes), but including $\ge$\,2\% energy contributions for all modes up to $\ell=7$. The fields of both components of \yy\ are predominantly poloidal (poloidal field contributes 71\% of the global field energy) and are split approximately equally between the axisymmetric ($m<\ell/2$) and non-axisymmetric ($m\ge\ell/2$) harmonic modes. The strictly axisymmetric (the sum of all $m=0$ modes) part of the magnetic geometries comprises 36--49\% of the field energy.

As can be clearly seen from the radial field distributions presented in Fig.~\ref{fig:zdi_maps}, the dipolar components of the magnetic fields of \yya\ and B are anti-aligned. The polarity of the field at the stellar North pole is predominantly positive for the primary and negative for the secondary.

We performed a series of 100 bootstrapping magnetic inversions to assess uncertainties of the ZDI results for \yy. In each such calculation the residuals of the fit to the LSD Stokes $V$ profiles were randomly reshuffled and added back to the synthetic spectra. The resulting data were subjected to the same tomographic analysis as the original observations. These calculations showed that the local standard deviation of the radial, meridional, and azimuthal magnetic field maps illustrated in Fig.~\ref{fig:zdi_maps} is 10--16~G on average. The maximum deviation is 19--27~G. The average magnetic quantities reported in Table~\ref{tbl:mag} are accurate to within 3--6~G. The uncertainty of the fractional energies of different harmonic components is 1.3--2.8\%.

\section{Zeeman Intensification Analysis}
\label{bi}

An analysis of magnetic broadening (for slowly rotating active stars) and intensification (more relevant for fast rotators such as the components of \yy) of absorption lines in the intensity spectra provides a powerful magnetic diagnostic method \citep[e.g.][]{basri:1992,reiners:2012}, which is complementary to, and in some cases more informative than, ZDI mapping of the global field topology. The latter technique can be applied to fields of essentially arbitrary strength and is particularly sensitive to field orientation. However, precisely due to this sensitivity, it yields a grossly underestimated field strength whenever stellar surface is peppered with small regions of opposite field polarity. In contrast, Zeeman broadening and intensification depend almost entirely on the field modulus. This prevents using these effects for inferring detailed surface magnetic field geometries. In addition, the field strength has to exceed $\sim$\,500 gauss to reliably disentangle magnetic effects from other processes impacting Stokes $I$ spectra. On the other hand, Zeeman intensification provides an unbiased measure of the total magnetic flux density, including both large and small-scale magnetic fields, and is free of the polarity cancellation problem inherent to most polarimetric diagnostic methods.

Our approach to measuring the total (unsigned) magnetic flux density in the components of \yy\ is based upon a direct spectrum synthesis modelling of the Zeeman effect in the intensity profiles of selected atomic absorption lines. 
Theoretical spectra required for this analysis were calculated with the help of the \textsc{Magnesyn} spectrum synthesis code described in \citet{shulyak:2017}. Considering the large rotational Doppler broadening of the spectra of \yya\ and B, it is impossible to derive unique field strength distributions comprised of multiple magnetic components as was done for slowly rotating M dwarfs in the latter study. Instead, we made use of a simpler two-component magnetic field model \citep[e.g.][]{kochukhov:2017c} to represent the observed line profiles. This model assumes that a fraction of the surface $f$ is covered by the field of strength $B$ and the rest of the surface $(1-f)$ is non-magnetic. The total surface-averaged magnetic field strength is then given by $\langle B_I \rangle = B\cdot f$. 

\begin{figure*}[!th]
\fifps{\hsize}{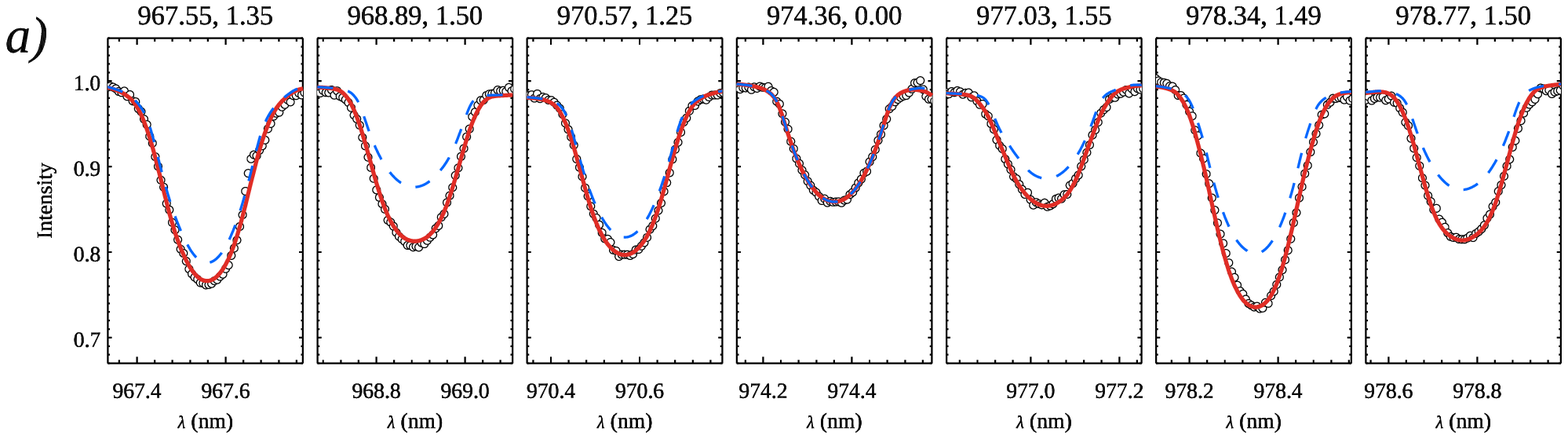}\vspace*{0.5cm}
\fifps{\hsize}{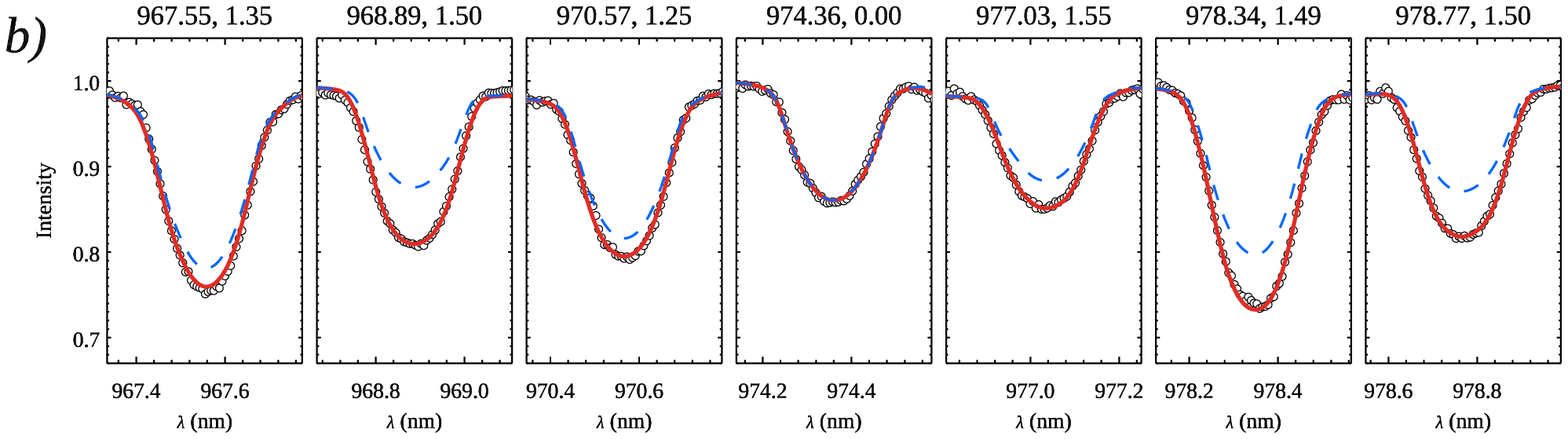}
\caption{
Zeeman broadening and intensification analysis for \yya\ (a) and \yyb\ (b) using seven Ti~{\sc i} spectral lines in the 967.4--978.8~nm wavelength interval. The time-averaged, disentangled observed spectra (open circles) are compared with the best-fitting magnetic calculations (thick solid red line) and with the corresponding non-magnetic spectrum synthesis (thin dashed blue line). The central wavelengths of the transitions and their effective Land\'e factors are indicated above each panel.
}
\label{fig:zb}
\end{figure*}

The thermodynamic structures of the magnetic and non-magnetic parts of the stellar surface were assumed to be the same and were approximated using models from the MARCS atmospheric grid. We assumed $T_{\rm eff}=3800$~K for both components as a compromise between the effective temperature $3820\pm100$~K given by \citet{torres:2002} and \citet{stassun:2016} and $\approx$\,$3770\pm100$~K determined by \citet{macdonald:2014}. The surface gravity $\log g=4.63$ was adopted according to the fundamental radius and mass of the mean component \citep{torres:2002}. The micro and macroturbulent velocities are reported to be on the order of 100~\ms\ by the three-dimensional hydrodynamic simulations of M dwarf atmospheres \citep{wende:2009}. Consequently, their exact choice has no impact on our analysis.

Large $v_{\rm e}\sin i$ of the \yy\ components makes it impossible to study Zeeman splitting or broadening in spectral lines. Instead, we rely on the effect of magnetic intensification to measure surface magnetic fields. For instance, in the case of a saturated spectral line, its equivalent width is proportional to the magnetic field intensity and depends on the separation and number of individual Zeeman components, the so-called Zeeman splitting pattern \citep{polarization:2004}. Therefore, at large $v_{\rm e}\sin i$ values intensities of individual spectral lines in a magnetic star will depend on their Zeeman patterns and will vary from line to line, making it possible to deduce magnetic field strength even for the most rapidly rotating stars.

The key to a successful application of the Zeeman intensification methodology is finding a set of unblended spectral lines with accurately known relative strengths and a different magnetic field response. As demonstrated by \citet{kochukhov:2017c} and \citet{shulyak:2017}, a group of Ti~{\sc i} lines in the 964--979~nm wavelength interval satisfies these requirements for stars with spectral types from early-M to about M6. There are 9 mostly unblended Ti~{\sc i} features, with one of them ($\lambda$ 974.36~nm) having null effective Land\'e factor and thus completely insensitive to a magnetic field. This line is useful for constraining non-magnetic parameters, such as $v_{\rm e}\sin i$ and Ti abundance. The remaining 8 lines have effective Land\'e factors from 1.00 to 1.55 and different Zeeman splitting patterns (computed using the line data from the VALD3 database), resulting in a different magnetic field response. All these Ti~{\sc i} lines belong to the same multiplet formed by the fine structure transitions between the a$^5$F and z$^5$F$^{\rm o}$ atomic terms. Their relative oscillator strengths are perfectly known, making it possible to accurately measure magnetic field strength by the spectrum synthesis fitting of the observed profiles of all or a subset of these Ti~{\sc i} lines. 

Telluric absorption usually severely contaminates the wavelength interval in question. However, our spectral disentangling procedure described in Sect.~\ref{sbfit} enabled an accurate telluric line removal and yielded clean, high SNR average profiles of Ti lines suitable for precise magnetic field measurements.

The disentangled time-averaged spectra of \yya\ and B derived above were corrected for the continuum dilution \citep[e.g.][]{folsom:2008} assuming equal luminosities of the components. The latter assumption appears to be adequate because, unlike the LSD profiles dominated by bluer lines, the disentangled spectra in the 963--985~nm region do not show a systematic line depth difference between the two components. The projected rotational velocity, Ti abundance and magnetic field parameters were optimised with the Levenberg-Marquardt  algorithm. We also allowed the code to vary continuum scaling factors for each spectral region (ensuring that fitting windows are sufficiently wide to include continuum on both sides of each line). This non-linear least-squares fitting was applied to different combinations of Ti~{\sc i} lines to assess stability of the derived magnetic field parameters. This analysis showed that fits to two of the Ti~{\sc i} lines, $\lambda$ 964.74 and 972.84 nm, yielded systematically higher residuals. The first of these features is located very close to the echelle order edge, is recorded at a lower SNR in all observations, and is shifted outside the wavelength coverage at some of the orbital phases. The second line has the smallest non-zero effective Land\'e factor for this Ti~{\sc i} multiplet and is blended by the Cr~{\sc i} $\lambda$ 973.03~nm line. These two Ti~{\sc i} lines were therefore excluded from the fit. The final field strength and filling factor values were derived from simultaneous analysis of 7 remaining Ti~{\sc i} lines.

The titanium abundance recovered for the \yy\ components in different fitting attempts did not differ by more than 0.04~dex from the solar value of $\log (N_{\rm Ti}/N_{\rm tot})=-7.09$ \citep{asplund:2009}. Considering that there are no physical reasons to expect a different chemical composition of the components of such a tight binary, we choose to adopt the solar Ti abundance for both components in the definitive analysis.

The final best-fitting model spectra are compared to observations in Fig.~\ref{fig:zb}. We also show the non-magnetic synthetic spectra computed with the same Ti abundance and $v_{\rm e}\sin i$. The effect of magnetic field is very prominent, especially for the Ti~{\sc i} $\lambda$ 968.89, 977.03, 978.34, 978.77~nm lines, which increase their equivalent width by up to 50\% compared to the non-magnetic spectrum. The resulting magnetic field parameters are given in Table~\ref{tbl:mag}. We obtained $\langle B_I \rangle = 3.44$~kG and $v_{\rm e}\sin i = 37.7$~\kms\ for \yya\ and $\langle B_I \rangle = 3.15$~kG, $v_{\rm e}\sin i = 37.9$~\kms\ for \yyb. The filling factors were estimated to be in the 0.75--0.86 range. The projected rotational velocities determined here are consistent with $v_{\rm e}\sin i$ that can be calculated from the rotational period, inclination angle, and the stellar radii determined in Sect.~\ref{mapi}.

The formal statistical errors calculated by the least-squares optimisation algorithm from the covariance matrix are often underestimating the true uncertainties. Rather than relying on these formal errors we made a conservative uncertainty estimate by calculating the standard deviations of $B$, $f$, and the corresponding $\langle B_I \rangle$ values inferred from separate modelling of 6 magnetically sensitive Ti~{\sc i} lines with the titanium abundance and $v_{\rm e}\sin i$ held constant. This analysis showed the line to line scatter of about 0.56 kG and 0.09 for the field strength and filling factor, respectively. On the other hand, their product remains the same to within 0.33 kG.

One may also be concerned that including a large number of magnetically sensitive lines biases the chi-square fit against the only magnetically insensitive feature, thereby leading to a degeneracy between magnetic field strength, $v_{\rm e}\sin i$ and Ti abundance. We tested this possibility by deriving the projected rotational velocity and titanium abundance from the non-magnetic line alone and then fitting the magnetic field parameters using magnetically sensitive lines. This did not change any of the results reported above, suggesting that the primary reason for the dependence of magnetic parameters on line selection is not the intrinsic degeneracy of our spectral fitting procedure but, more likely, an unrecognised line blending, imperfect removal of telluric features and, possibly, variation of the spectrograph's instrumental profile across the echelle order.

According to the error analysis performed by \citet{shulyak:2017}, variation of $T_{\rm eff}$ by 100~K leads to $\le$\,0.05~kG change of $\langle B_I \rangle$ inferred for early M dwarfs. Thus, the Zeeman intensification results presented here are not significantly affected by $\approx$100~K uncertainty of the mean $T_{\rm eff}$ of the \yy\ components. Likewise, our neglect of the possible 90~K difference of the component temperatures (see Sect.~\ref{mapi}) cannot lead to the 0.3~kG stronger mean field obtained for \yy~A.

Finally, we have attempted to investigate rotational profile variability of individual Ti lines modelled in this section. Due to limited SNR of the observed data, we could not detect any profile variability pattern similar to the one seen in the LSD profiles constructed from bluer lines. We also did not detect any differential variability of the Ti~{\sc i} lines with large and small Land\'e factors. These results show that there are no significant non-uniformities in the surface distribution of the small-scale magnetic field in either of the \yy\ components and that these field strength inhomogeneities do not contribute to the Stokes $I$ profile variation. This justifies approximating the entire stellar surface with a single magnetic field strength distribution.

\section{Summary and Discussion}
\label{disc}

This study presented a comprehensive analysis of the surface magnetic fields of the components of the key benchmark early-M eclipsing binary system \yy. We used high quality archival spectropolarimetric observations to obtain precise radial velocity measurements and revise spectroscopic orbital elements, leading to an improved estimate of the component masses. The spectral disentangling technique was applied to separate contributions of the components in the mean intensity line profiles and in individual wavelength regions. We applied the tomographic imaging techniques of DI and ZDI to reconstruct maps of dark spots from variability of the mean intensity profiles and global magnetic field topologies from rotational modulation of the least-squares deconvolved circular polarisation signatures. We then carried out a complementary analysis of the differential Zeeman intensification of individual magnetically sensitive atomic lines and inferred the total mean field strength at the surfaces of both \yy\ components.

\subsection{The Global and Total Magnetic Fields of M Dwarfs}

The outcome of our ZDI investigation indicates that both stars in the \yy\ system possess non-axisymmetric, predominantly poloidal global magnetic field geometries. The dipolar components of these field structures are approximately anti-aligned, hinting at the magnetospheric interaction in the system. The mean global field strengths are found to be 205--260~G, with the primary exhibiting about 27\% stronger field than the secondary. This difference in the global field characteristics is consistent with the higher X-ray activity of the primary \citep{hussain:2012}. At the same time, the Zeeman intensification analysis yields the total mean field strengths of 3.44 and 3.15 kG for \yya\ and B, respectively. The primary is still found to be more magnetised than the secondary, albeit only marginally so considering $\sim$\,0.3~kG errors of these field strength measurements. This is reminiscent of the behaviour of the components of the fully convective M dwarf binary GJ\,65 \citep{kochukhov:2017c}, which also have nearly identical fundamental parameters, rotation periods, and show comparable total field strengths while exhibiting vastly different global field geometries.

The much stronger field obtained with the Stokes $I$ modelling indicates that the surface magnetic field topologies of \yya\ and B are dominated by small-scale, tangled field structures, which are not detectable in Stokes $V$. Our study suggests that the ratio of the global to total field strength is 6.5--7.6\% or, equivalently, that only 0.4--0.6\% of the total magnetic field energy is contained in the global field. \citet{reiners:2009} compared the global and total field strengths for 6 M dwarfs, showing that their ratio drops from $\sim$\,15\% in fully convective M dwarfs to only $\sim$\,0.5\% in early-M stars. Our measurements for the components of \yy\ appear to agree with this trend.

\begin{figure}[!t]
\fifps{8.5cm}{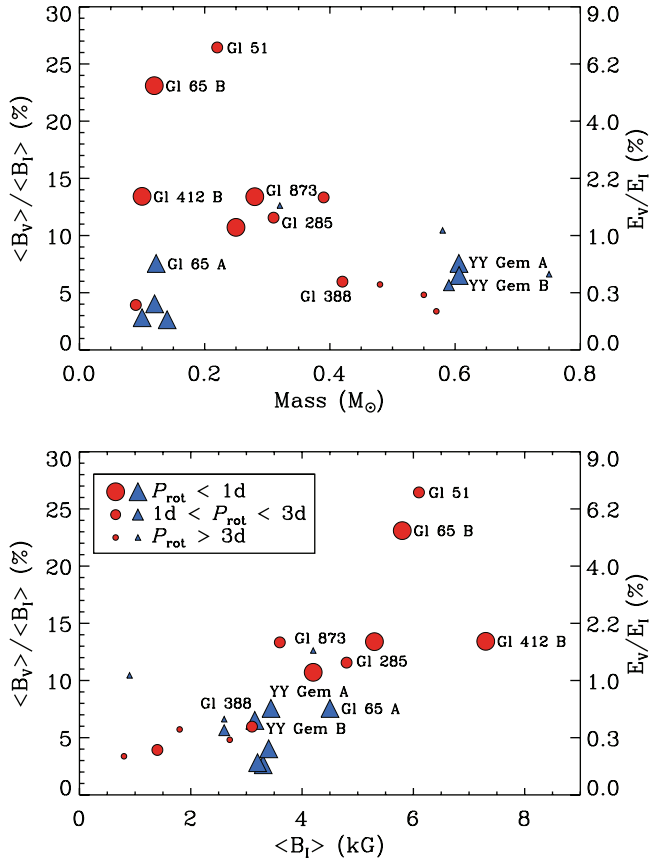}
\caption{
Ratio of the mean field strengths derived from Stokes $V$ and Stokes $I$ analyses and the corresponding magnetic field energy ratio as a function of stellar mass (upper panel) and the field strength obtained from Stokes $I$ (lower panel). Red circles show stars with predominantly dipolar, axisymmetric fields. Blue triangles correspond to stars with non-axisymmetric, multipolar fields. The symbol size reflects rotation period, as indicated by the legend in the lower panel.
}
\label{fig:trend}
\end{figure}

The sample of M dwarfs with accurate Zeeman broadening field strength measurements has been significantly expanded since the work by \citet{reiners:2009}. It is therefore pertinent to re-examine the relation of global and total magnetic fields using new data. To this end, we combined information on the mean global field strengths $\langle B_V \rangle$ reported by ZDI studies of M dwarfs \citep{morin:2008,donati:2008,morin:2010,kochukhov:2017c} with the total field strengths $\langle B_I \rangle$ determined for the same stars from Stokes $I$ analysis \citep{shulyak:2017}. Including results obtained here for the \yy\ components yields a sample of 22 individual M dwarfs with spectral types ranging from M0 to M6.5. Distribution of the field strength and field energy ratios as a function of stellar mass and total mean field strength is presented in Fig.~\ref{fig:trend}. The symbol size in this figure reflects the stellar rotation period while the symbol style corresponds to the classification of global magnetic field topologies of M dwarfs into axisymmetric, mainly dipolar, and non-axisymmetric, primarily multipolar, configurations according to \citet{shulyak:2017}. In this framework, the components of \yy\ are grouped with multipolar stars due to the large contribution of non-dipolar and non-axisymmetric harmonic modes.

The panels of Fig.~\ref{fig:trend} suggest that a systematic trend of the fraction of magnetic  energy contained in the global field component with stellar mass and total field strength is limited to a subgroup of stars with predominantly dipolar, axisymmetric global fields. Multipolar field stars do not show any obvious trends, although their position in the $\langle B_V \rangle / \langle B_I \rangle$ vs. $\langle B_I \rangle$ plane agrees with the trend shown by M dwarfs with the global dipolar fields. The upper panel of Fig.~\ref{fig:trend} reveals that the two very active mid-M dwarfs Gl~65~B (UV~Ceti) and Gl~51 have an unusually high global to total field strength ratio of $\sim$\,25\% (i.e. $\sim$\,6\% of the total field energy is stored in the global field component). One can therefore conclude that the fields in these stars have a weaker, though still dominant, local tangled contribution. Interestingly, Gl~412~B (WX UMa), which possesses the strongest known field in an M dwarf, deviates significantly from the general field complexity trends. 

The main message of Fig.~\ref{fig:trend} is that only a small fraction of the total magnetic field energy is visible in circular polarisation and can thus be recovered by ZDI. This fraction varies from $>0.1$ to 7.0\%, depending on the stellar parameters, geometry of the global field and the total field strength. This situation is very different from the behaviour of massive, early-type stars with pure dipole-like fossil fields for which simple global magnetic field models are able to simultaneously reproduce both polarisation and intensity magnetic observables \citep[e.g.][]{landstreet:2000,kochukhov:2015}. 

A complete characterisation of an M dwarf surface magnetic field therefore requires a combined investigation of the Zeeman effect in intensity spectra and polarisation profile modelling with ZDI. The latter analysis alone cannot provide information about such key magnetic field characteristic as the total field strength. For this reason ZDI results should be used with extreme caution for inferences regarding underlying dynamo mechanism or its dependence on the stellar mass and rotation. How exactly the global (organised) and the local (tangled) magnetic field components combine at the stellar surface remains an open question. Some ideas of how such composite field geometries may look like have been proposed based on empirical arguments \citep{lang:2014} and \textit{ab initio} three-dimensional MHD simulations \citep{yadav:2015}. However, since no comprehensive polarised spectrum synthesis calculations were carried out for either of these modelling frameworks, it is not known how they would fare when confronted with real observations.

\subsection{Testing Predictions of Magnetoconvection Stellar Structure Models of Low-Mass Stars}

Detailed characterisation of the surface magnetic field of the components of \yy\ provided by our study puts us in the position to directly test predictions of the magnetic stellar interior structure models developed for this system with the aim to explain the inflated radii. In one of such studies \citet{feiden:2013} were able to reproduce the observed mean radius of the \yy\ components by invoking stabilisation of the interior convection by a magnetic field. They found that the surface magnetic field strength required by their best-fitting models lies in the range from 4.0 to 4.5~kG, which is only 20--40\% higher than the total surface field strength $\langle B_I \rangle$ measured here using Zeeman intensification. 

An alternative approach to incorporating effects of a magnetic field in one-dimensional interior models of low-mass stars was developed by \citet{mullan:2001} and \citet{macdonald:2014}. The latter paper presented an in-depth modelling of the \yy\ system and showed that its properties can be reproduced with various interior magnetic field profiles corresponding to 240--420~G vertical field on the stellar surface. \citet{macdonald:2017} have subsequently revised this prediction to 490--550~G.

There is some confusion in the literature regarding the specific magnetic field observable which should be compared with the mean vertical field discussed in these theoretical papers. \citet{macdonald:2017a} argued that their vertical surface field parameter corresponds to the global field inferred by ZDI while the surface field in \citet{feiden:2013} models should be compared to $\langle B_I \rangle$ measurements. On the other hand, in their recent study of the wide M dwarf binary GJ~65 \citet{macdonald:2018} compared their model predictions with $\langle B_I \rangle$ rather than with any field characteristics obtained from circular polarisation modelling.

Considering our ZDI results, the mean vertical field corresponds to the surface-averaged unsigned radial field $\langle |B_{\rm r}| \rangle$ reported in Table~\ref{tbl:mag}. Our $\langle |B_{\rm r}| \rangle$\,=\,98--168~G is a factor of 2.9--5.6 lower than the vertical surface fields suggested by \citet{macdonald:2017}. If, instead, we choose to compare their predictions with our $\langle B_I \rangle$ measurements, the theoretically estimated field should be multiplied by $\sqrt{3}$ assuming isotropic field vector orientation. This yields a total surface field strength of 850--950~G, which is 3.3--4.0 times lower than $\langle B_I \rangle$ determined in our paper.

To summarise, theoretical models by \citet{feiden:2013} appear to predict a more realistic surface field strength for the components of \yy\ than the one found in the approach by \citet{macdonald:2014,macdonald:2017}. However, this conclusion should be taken with caution given the ambiguity of relating the vertical surface field parameter of the latter theoretical calculations with observations.

\subsection{Estimating Magnetic Flux from X-ray Luminosity}

In the absence of direct constraints on the surface magnetic properties many studies resort to an approximate estimate of the mean magnetic field strength. The X-ray emission is often considered to be a particularly useful magnetic proxy. \citet{pevtsov:2003} demonstrated that the total unsigned magnetic flux measured for solar magnetic regions at different spatial scales shows a power law dependence on the X-ray spectral radiance. The few $\langle B_I \rangle$ measurements reported at that time for active late-type stars fell on the extrapolation of this relation to higher magnetic flux values. \citet{feiden:2013} recalibrated this relation using Stokes $I$ field strength measurements of low-mass dwarfs. Both \citet{feiden:2013} and \citet{macdonald:2014} attempted to verify their theoretically predicted surface field strengths with indirect field strength estimates using the X-ray emission of \yy. This exercise was, however, hampered by the contamination of the ROSAT X-ray measurements of \yy\ \citep{voges:1999} by the nearby early-type companions Castor A and B. 

More recent Chandra observations of \yy\ \citep{hussain:2012} provided images with a high angular resolution, allowing one to separate the low-mass binary from the A-type stars. The combined quiescent X-ray luminosity of \yy\ is reported to be 3--3.5$\times10^{29}$~erg s$^{-1}$. Assuming equal luminosities of the components and using their fundamental radii yields $\langle B_I \rangle$\,=\,770--820~G with the calibration by \citet{feiden:2013}. This is significantly smaller than the observed $\langle B_I \rangle$ obtained in our study, indicating that the magnetic flux vs. X-ray luminosity relation may be appropriate only for large stellar samples but is unlikely to be precise enough to provide a meaningful field estimate for individual stars.

Another problematic issue with applying this relation to active stars, apparently entirely overlooked in the literature, is that for the Sun the total magnetic flux is defined as a surface integral of the unsigned vertical field component $|B_z|$ \citep{fisher:1998}. Therefore, an extension of this relation to stars requires introducing a correction factor, dependent on the surface field geometry (e.g. $\sqrt{3}$ for isotropic field, 1.38 for a dipolar field topology, a value $\gg1$ for a field configuration dominated by toroidal component), to account for the difference between $\langle |B_z| \rangle$ and $\langle B_I \rangle$. This correction would systematically reduce the observed surface magnetic fluxes for stars employed to calibrate the relation with X-ray luminosity. Since the ratio of poloidal and toroidal magnetic energy is known to change systematically with stellar parameters and mean magnetic field strength \citep{petit:2008a,see:2015}, it is plausible to expect that this reduction is not universal but varies significantly from one star to another.

\subsection{Spot Filling Factor for the Components of \yy}

The tomographic mapping of star spots carried out here for \yy\ provided new detailed information on the characteristics of non-uniform brightness distributions on these stars. We found that both components possess low-contrast, relatively small-scale surface features, concentrated at low latitudes. If we interpret our Doppler imaging maps in terms of the fractional coverage by completely dark spots, we infer a spot coverage of $<1$\% for both components. This can be compared with 3\% spot coverage obtained with the help of photometric light curve modelling assuming few large circular spots located at fixed latitudes \citep{torres:2002}.

Formation of cool spots on the surfaces of magnetically active late-type stars was proposed as an alternative explanation of their increased radii \citep{chabrier:2007}. \citet{macdonald:2014} applied this hypothesis to \yy, finding that dark spots must cover more than 40\% of the surface area of each component in order to match the observed radii. Likewise, \citet{jackson:2014} estimated that the observed radii inflation of young stars in open clusters and low-mass eclipsing binaries, including \yy, can be reproduced by stellar models with star spot filling factors in the 35--51\% range. This assessment corresponds to completely dark spots. The required spot coverage must be even larger for a finite spot-to-photosphere brightness contrast. Such extreme surface inhomogeneities are clearly not observed for \yy. Our detailed line profile analysis also shows no evidence of polar spots which could have been missed by previous photometric studies. In general, cool polar caps appear to be uncommon on rapidly rotating early to mid-M dwarfs \citep{barnes:2004a,barnes:2017,morin:2008a}. Therefore, we conclude that our results do not support any theoretical explanation of the properties of \yy\ which involves postulating substantial cool spot coverage. On the other hand, we cannot exclude existence of numerous, homogeneously distributed small spots with sizes below the resolution limit ($\approx$10\degr) of our DI analysis. This scenario can be tested with very high precision photometric monitoring of eclipses in the \yy\ system. The required observational data may be provided by the TESS mission \citep{ricker:2016}, which is scheduled to observe \yy\ in January 2020.

\acknowledgements
OK acknowledges support by the Knut and Alice Wallenberg Foundation (project grant ``The New Milky Way''), the Swedish Research Council (project 621-2014-5720), and the Swedish National Space Board (projects 185/14, 137/17).
DS acknowledges support by DFG in the framework of the project ``CRC 963: Astrophysical Flow Instabilities and Turbulence'' (projects A16 and A17).
The authors thank Ansgar Reiners for fruitful discussions.

This paper is based on observations obtained at the CFHT which is operated by the National Research Council of Canada, the Institut National des Sciences de l'Univers (INSU) of the Centre National de la Recherche Scientifique (CNRS) of France, and the University of Hawaii.
Archival data were retrieved using facilities of the Canadian Astronomy Data Centre operated by the National Research Council of Canada with the support of the Canadian Space Agency. 

This work has made use of the VALD database, operated at Uppsala University, the Institute of Astronomy RAS in Moscow, and the University of Vienna.

\facility{CFHT (ESPaDOnS spectropolarimeter)}

\bibliographystyle{aasjournal}
\bibliography{astro_papers}

\end{document}